\documentclass[sigconf]{acmart}

\usepackage{threeparttable}
\usepackage[utf8]{inputenc} 
\usepackage{nicefrac}       
\usepackage{microtype}
\usepackage{soul}
\usepackage{url}
\usepackage{algorithm}
\usepackage{algorithmic}
\usepackage{multirow}
\usepackage{subcaption}
\usepackage{bm}
\usepackage{tablefootnote}
\usepackage{lipsum}
\usepackage{tcolorbox}
\usepackage{xspace}
\usepackage{stfloats}
\usepackage{marvosym}
\usepackage{pifont}
\usepackage{subcaption}

\usepackage{booktabs}

\setcopyright{acmcopyright}
\copyrightyear{2018}
\acmYear{2018}
\acmDOI{XXXXXXX.XXXXXXX}
\acmConference[Conference acronym 'XX]{Make sure to enter the correct
  conference title from your rights confirmation emai}{June 03--05,
  2018}{Woodstock, NY}
\acmPrice{15.00}
\acmISBN{978-1-4503-XXXX-X/18/06}

\begin{document}

\title{RankUp: Towards High-rank Representations for Large Scale Advertising Recommender Systems}

\author{Jin Chen$^*$, Shangyu Zhang$^*$, Bin Hu$^*$, Chao Zhou$^{* \dagger}$, Junwei Pan, Gengsheng Xue, Wentao Ning, Gengyu Weng, Wang Zheng, Shaohua Liu, Zeen Xu, Chengyuan Mai, Shijie Quan, Tingyu Jiang, Lifeng Wang, Shudong Huang, Chengguo Yin, Haijie Gu, Jie Jiang}
\email{elinjinchen, vitosyzhang, glennbinhu, derekczhou, jonaspan, eddiegsxue, wentaoning, greyweng, wangzheng@tencent.com}
\email{
jimmyshliu, flashxu, henrymai, justinquan, travisjiang, fandywang, ericdhuang, turingyin, jerrickgu, zeus@tencent.com}
\affiliation{
    \country{Tencent Inc.}
    % \country{Shenzhen, China}
}

\renewcommand{\shortauthors}{Jin Chen, Shangyu Zhang, Bin Hu, Chao Zhou and et al.}

\begin{abstract}
The scaling laws for recommender systems have been increasingly validated, where MetaFormer-based architectures consistently benefit from increased model depth, hidden dimensionality, and user behavior sequence length.
However, whether representation capacity scales proportionally with parameter growth remains unexplored.
Prior studies on RankMixer reveal that the effective rank of token representations exhibits a damped oscillatory trajectory across layers, failing to increase consistently with depth and even degrading in deeper layers.
Motivated by this observation, we propose \textbf{RankUp}, an architecture designed to mitigate representation collapse and enhance expressive capacity through randomized permutation splitting over sparse features, a multi-embedding paradigm, global token integration and crossed pretrained embedding tokens.
RankUp has been fully deployed in large-scale production across Weixin Video Accounts, Official Accounts and Moments, yielding GMV improvements of 3.41\%, 4.81\% and 2.12\%, respectively.

\end{abstract}

\keywords{Recommender Systems, Dimensional Collapse}

\thanks{$^*$Equal contribution.}
\thanks{$\dagger$Corresponding author.}
\maketitle

\section{Introduction}
Recommendation architectures have evolved from manual feature interaction modeling toward deep representation learning, with Transformer-based structures emerging as the dominant paradigm in modern industrial recommender systems.
Inspired by the scaling laws observed in large language models (LLMs)~\cite{achiam2023gpt}, recent industrial recommendation models — including AutoInt~\cite{songweiping2019AutoInt}, Hiformer~\cite{gui2023hiformer}, Wukong~\cite{zhang2024wukong}, Interformer~\cite{zeng2025interformer}, RankMixer~\cite{zhu2025rankmixer}, TokenMixer-Large~\cite{jiang2026tokenmixer} and Mixformer~\cite{huang2026mixformer} — have increasingly converged toward the MetaFormer~\cite{yu2022metaformer} paradigm.
These architectures consistently demonstrate improved ranking performance through scaling model depth, hidden dimensionality, and user behavior sequence length, validating the scaling laws for recommender systems.
Despite these empirical gains, an important question remains largely unexplored: does representation capacity improve with parameter scaling?
Scaling model size and enhancing representation capacity are not equivalent. Existing approaches largely assume that larger models yield better representations, yet they overlook structural constraints that inherently limit representational expressiveness. Consequently, scaling alone often leads to diminishing returns, where additional parameters contribute little to effective capacity and may even exacerbate representation collapse in deep architectures.

Representation capacity plays a critical role in the scalability of recommendation models, as limited expressive power often leads to embedding collapse, where learned representations concentrate in low-dimensional subspaces, reducing the model's ability to differentiate users and items and thereby limiting ranking performance.
To characterize this phenomenon, prior studies commonly measure representation capacity through spectral metrics such as effective rank~\cite{roy2007effective} and fractional effective rank~\cite{wang2025attention}, which quantify the diversity and utilization of learned latent representations.
However, recent analyses~\cite{guo2026rankelastor} empirically reveal that deeper recommendation architectures do not necessarily exhibit increased representation capacity: as model depth increases, the effective rank of token representations follows a damped oscillatory trajectory across layers, failing to grow monotonically and often deteriorating in deeper layers.  
Theoretically, this behavior stems from the intrinsic limitations of existing RankMixer, where token mixing provides only bounded rank expansion while FFN exhibit rank-contractive behavior, jointly preventing sustained representation growth as depth increases.

While one could attempt to improve representation capacity by designing more sophisticated token mixing operations, such as Self-Attention~\cite{vaswani2017Transformer}, Full-Mix~\cite{guo2026rankelastor}, or Unimixer~\cite{ha2026unimixer}, such approaches primarily enhance feature interactions within an already constrained representation space and thus offer limited ability to prevent representation collapse in deep layers. In contrast, enhancing the expressive diversity of the latent space directly increases the capacity of the representation itself, enabling the model to preserve richer and more distinguishable token representations throughout depth. Motivated by this insight, we propose ~\textbf{RankUp}, an architecture that improves latent space diversity through:
Randomized Permutation Splitting reduces correlation and collinearity between tokens compared to semantic grouping, generating richer combinations; The Multi-embedding Representation Paradigm expands the foundational degrees of freedom in latent space; Global Token Integration enables each token to interact with global contextual information during token mixing; Cross Integration of Pre-trained Embeddings incorporates knowledge from other domains, scenarios, or tasks to enrich the latent space with external priors. Finally, Task-Specific Token Decoupling mitigates gradient interference in multi-objective settings, ensuring that the expanded parameter space is effectively utilized. Collectively, these mechanisms enhance the flexibility and diversity of latent representations, alleviating embedding collapse and supporting stronger representation capacity in deep layers.

We evaluate RankUp on large-scale production datasets from Tencent's advertising platforms, focusing on Click Conversion Rate (CVR) optimization. Extensive ablation studies show that each of the proposed mechanisms contributes to mitigating embedding collapse and enhancing latent representation diversity. These improvements translate to stronger online performance, with real-time AUC gains over the production baseline. Online A/B testing on 20\% of traffic further demonstrates significant uplift, and full deployment across Weixin Video Accounts, Moments, and Official Accounts results in GMV increases of 3.41\%, 4.81\%, and 2.12\%, respectively.
\section{Preliminaries}
\subsection{Formulation}
We consider a large-scale recommendation system with heterogeneous inputs, including sparse categorical IDs, dense numerical vectors, and user behavior sequences. For $M$ sparse features $\mathcal{F} = \{f_1, \dots, f_M\}$, each $f_i$ is mapped to an embedding $\mathbf{e}_i \in \mathbb{R}^{d_i}$. 
Due to the hundreds of sparse features, a splitting mechanism is typically employed to partition the high-dimensional feature space into a sequence of token representations. One prevalent approach is Autosplit~\cite{zhu2025rankmixer}, which partitions the concatenated vector $\mathbf{e}_{\text{input}} = [\mathbf{e}_1 ; \mathbf{e}_2; ...; \mathbf{e}_M]$ into $T$ equidistant segments of length $d_s$. Formally, the $i$-th token $\mathbf{x}_i$ is derived as: $$ \mathbf{x}_i = \text{Proj}_{i}(\mathbf{e}_{\text{input}}[ d_s \cdot (i - 1): d_s\cdot i]) $$ where the function $\text{Proj}$ maps the split embedding into $D$ dimension. Alternatively, a semantic-based approach leverages domain expertise to pre-define groups of features, where related features $\mathcal{F}_i \subset \mathcal{F}$ are aggregated based on domain expertise. The $i$-th token representation $\mathbf{x}_i$ is then formulated by projecting the concatenated embeddings within the group: $$ \mathbf{x}_i = \text{Proj}(\text{Concat}(\{\mathbf{e}_j \mid f_j \in \mathcal{F}_i\})) $$ This ensures that features with intrinsic semantic correlations are fused into an unified token prior to high-order interaction modeling.
Other auxiliary data (e.g., dense vectors, sequential representations) are similarly projected into the $D$-dimensional latent space. These tokens form the initial representation matrix $\mathbf{H}_0 \in \mathbb{R}^{T \times D}$.

Modern industrial models typically follow the MetaFormer~\cite{yu2022metaformer} paradigm, which decouples token-wise interactions from channel-wise transformations. The backbone consists of $L$ stacked layers, each composed of a Token Mixer and a Per-token FFN:$$ \mathbf{H}'_l = \text{TokenMixer}(\text{LN}(\mathbf{H}_{l-1})) + \mathbf{H}_{l-1} $$$$ \mathbf{h}_{l,i} = \text{FFN}(\text{LN}(\mathbf{h}'_{l,i})) + \mathbf{h}'_{l,i}, \quad i=1, \dots, T $$where $\text{LN}(\cdot)$ denotes Layer Normalization. In RankMixer, the Token Mixer employs a multi-head token mixing strategy: each token embedding is split into multiple heads, and the sub-vectors of all tokens are recombined across the sequence to facilitate global cross-token feature interaction. This operation is parameter-free, allowing efficient information exchange while maintaining high parallelism. The per-token FFN, in contrast, applies independent transformations to each token, preserving permutation equivalence and enhancing expressive capacity along the channel dimension.

\subsection{Representation Rank Deficiency}
To quantitatively analyze the utilization of the latent space in recommendation models, we consider the Effective Rank of the hidden representation matrix $\mathbf{H}_l \in \mathbb{R}^{T \times D}$ at each layer $l$. Unlike the standard algebraic rank, which is highly sensitive to infinitesimal noise, the effective rank provides a robust measure of the internal dimensionality or the distribution of information across the $D$ embedding dimensions.

Formally, let $\sigma_1, \dots, \sigma_k$ denote the singular values of $\mathbf{H}^l$ obtained via Singular Value Decomposition (SVD), where $k = \min(T, D)$. The effective rank is defined as the exponential of the Shannon entropy of the normalized singular value distribution $p_i = \sigma_i / \sum_{j=1}^k \sigma_j$:
$$\text{erank}(\mathbf{H}_l) = \exp \left( -\sum_{i=1}^k p_i \ln p_i \right)$$
This metric ranges from $1$ to $k$, where a value close to $1$ indicates that the representation has collapsed into a single dominant dimension, while a value approaching $k$ implies that information is uniformly distributed across all orthogonal dimensions.

Empirical observations from prior work~\cite{guo2026rankelastor} reveal that in deep recommendation architectures such as RankMixer, the effective rank of token representations does not monotonically increase with depth. Instead, it exhibits a damped oscillatory trajectory across layers: after initial expansion, the rank often stagnates or even decreases in deeper layers. This pattern indicates that despite scaling model depth and parameters, the latent space is not fully exploited, and the representations tend to collapse into low-dimensional subspaces. Such observations point to a fundamental limitation of existing token mixing and per-token FFN designs. Theoretically, token mixers can only provide bounded rank expansion across tokens, while per-token FFNs exhibit rank-contractive behavior along the channel dimension. Together, these intrinsic properties constrain the latent space, preventing sustained growth of representation diversity across layers. This theoretical insight motivates the need for architectural interventions that explicitly enhance the expressive capacity of the latent space.
\section{Rank-Up Method}

\subsection{Overall Framework}
Motivated by the limitations of existing token mixing and per-token FFN designs—namely, bounded rank expansion across tokens and rank contraction along channels—RankUp enhances latent space diversity by reducing token correlations, expanding embedding degrees of freedom, integrating global context, incorporating cross-domain priors, and decoupling task-specific representations.

\begin{figure*}
    \centering
    \includegraphics[width=0.95\linewidth]{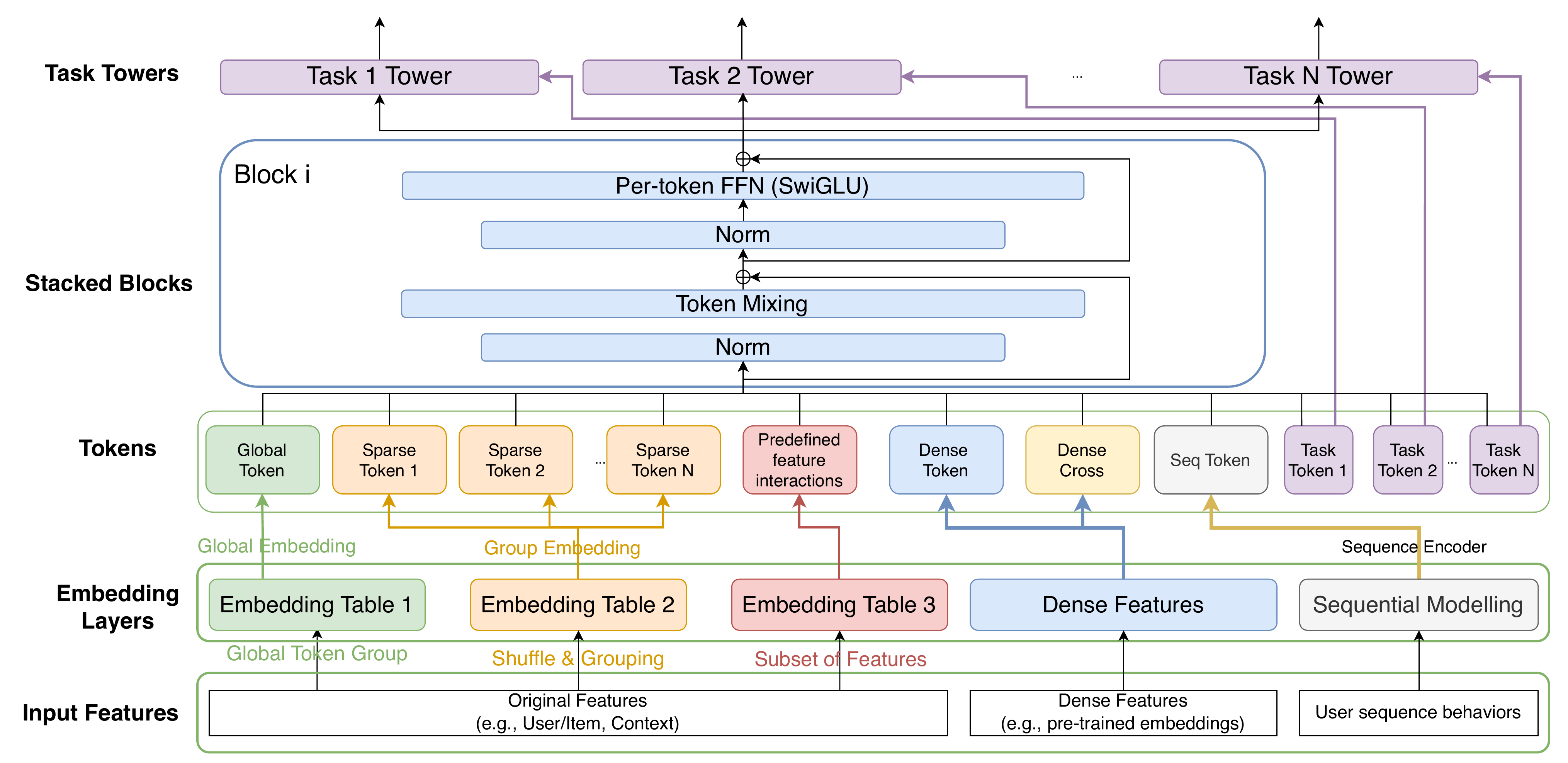}
    \caption{Overall Framework of RankUp}
    \label{fig:overall_framework}
\end{figure*}

Building on the MetaFormer backbone (Section 2), RankUp addresses latent space under-utilization, aiming to preserve and enrich the effective rank of token representations. Rather than relying solely on more sophisticated token mixing, it enhances the expressive diversity of the latent space, giving the model greater flexibility to maintain rich and distinguishable representations (Figure~\ref{fig:overall_framework}).
Randomized Permutation Splitting reduces correlations and collinearity between tokens, generating richer combinatorial interactions. A Multi-embedding Representation Paradigm expands the degrees of freedom of initial token embeddings, increasing latent diversity. A Global Token captures holistic context, enabling interactions across all tokens. Cross Integration of Pre-trained Embeddings injects knowledge from other domains, and Task-Specific Token Decoupling mitigates gradient interference in multi-objective settings. Together, these mechanisms enhance latent flexibility, alleviate embedding collapse, and support stronger representation capacity in deeper layers.

To ensure stable optimization, we adopt Pre-Layer Normalization (PreNorm)~\cite{xiong2020layer,wang2024deepnet} and SwiGLU activations~\cite{shazeer2020glu, chowdhery2023palm} within the per-token FFN. PreNorm stabilizes gradient flow in deep layers, facilitating more reliable training. The gating mechanism within SwiGLU enhances the modeling of complex feature interactions.

\subsection{Randomized Permutation Splitting}
Traditional industrial rankers divide the $M$ sparse features $\mathcal{F}$ into $T$ tokens according to predefined groupings~\cite{chenghengtez2016wide_and_deep,covington2016Youtube}. As described in Section 2.1, these approaches either perform equidistant segmentation of a fixed concatenation $\mathbf{e}_{\text{all}}$ (Autosplit) or assign features according to predefined semantic sets $\mathcal{F}_t$ (Semantic Grouping). Although these predefined groupings introduce a structured inductive bias, they often incur information redundancy, as highly collinear features clustered within the same token reduce the informational entropy~\cite{roy2007effective}. To address this limitation, we propose Randomized Permutation Splitting, which decouples feature grouping from fixed structural or semantic priors.
Formally, we define a stochastic permutation operator $\sigma$ that randomly shuffles the indices of the sparse feature set $\mathcal{F}$:
$$\mathcal{F}_\sigma = \{f_{\sigma(1)}, \dots, f_{\sigma(M)}\}$$ After applying Randomized Permutation Splitting, features within each group are vectorized and concatenated before projection.
By randomly distributing highly correlated features across tokens, Randomized Permutation Splitting reduces inter-token correlation. This decorrelation expands the geometric basis of the initial representation matrix $\mathbf{H}_0$, encouraging the model to learn a high-rank latent manifold and mitigating representation collapse in deeper layers.

\subsection{Multi-embedding Representation Paradigm}
Conventional industrial rankers typically map sparse features to a latent space using a single embedding table $\psi: \mathcal{F} \to \mathbb{R}^d$. While computationally efficient, this Single-embedding paradigm imposes a fixed, low-dimensional constraint on the input manifold, often limiting the informational diversity required by downstream token mixers. This constraint induces a representation bottleneck, forcing high-order interaction layers to operate on a compressed feature basis and increasing the risk of representation collapse. To address this issue, we adopt a Multi-embedding Paradigm~\cite{pan2024ads, guo2023embedding} within the token set. Rather than using a single projection, we employ $K$ independent embedding tables ${\psi_1, \dots, \psi_K}$ to represent input features. For a feature $f_j$, its representation becomes a tuple of embeddings drawn from different tables: $\mathbf{e}_{j} = \{ \psi_k(f_j) \mid \psi_k \in \mathcal{K}_j \}$, where $\mathcal{K}_j$ denotes the subset of tables assigned to $f_j$. This redundant mapping allows the same categorical signal to be represented from multiple geometric perspectives, providing a more granular initialization for subsequent token formation.

Leveraging this multi-embedding architecture increases the diversity of the input representation matrix $\mathbf{H}_0$. This high-rank initialization alleviates the early-stage bottleneck common in single-embedding systems, enabling the model to better capture rare signals in sparse scenarios.

\subsection{Global Token Integration}
In typical MetaFormer-based architectures, each token represents information from a local subset of features.
To provide a holistic view of the input for interaction, we introduce a Global Token $\mathbf{g}$ alongside the local tokens. Unlike local tokens, which encode intra-group interactions, the Global Token $\mathbf{g}$ aggregates information from all features ${f_1, \dots, f_M}$ via an aggregation function $A$:$$\mathbf{g} = A(f_1, f_2,..., f_M) = \text{func}\left( \text{Pool}(\{\text{Embed}(f_i)\}_{i=1}^M) \right)$$
The function $\text{func}$ can be implemented as a Multi-Layer Perceptron (MLP) or more sophisticated cross-interaction modules, such as FM~\cite{koren2009MF,panjunwei2018FwFM,juanyuchin2016FFM} or DCNv2~\cite{wangruoxi2021DCNv2}, to capture global dependencies. This Global Token is then appended to the token sequence $\mathbf{H}^{(0)} = [\mathbf{g}, \mathbf{e}_1, \dots, \mathbf{e}_T]$, providing global contextual information for subsequent token interactions.

\subsection{Cross Integration of Pre-trained Embeddings}
In modern industrial ranking systems, pre-trained user and item embeddings are widely adopted as a fundamental component. These embeddings are typically learned from large-scale historical interaction data using two-tower retrieval models, capturing coarse-grained semantic similarity in a dense latent space~\cite{zhang2024scaling, zhang2025large}. However, such embeddings are primarily optimized under distance-based objectives, focusing on global similarity rather than fine-grained feature interactions required in ranking scenarios. In contrast, ranking models fundamentally rely on explicit feature interactions to capture higher-order dependencies among heterogeneous signals. As a result, directly using pre-trained embeddings in a shallow manner (e.g., concatenation or linear projection) fails to fully expose their interaction structure to downstream layers.
To bridge this gap, we introduce a cross integration mechanism that explicitly injects interaction priors into the representation space. Given the user and item embeddings $\mathbf{z}_{ue}$ and $\mathbf{z}_{ie}$, we compute their interaction via element-wise product:
$$\mathbf{e}_{cross} = \text{Proj} \left(  \mathbf{z}_{ue} \odot \mathbf{z}_{ie}  \right)$$
which can be viewed as a soft feature-level interaction that aligns with the inductive bias of factorization-based ranking models.
This token is appended to the input sequence $\mathbf{H}_0$, explicitly injecting interaction-aware prior knowledge from pre-trained models into the token-based ranking architecture, complementing sparse feature interactions in downstream layers.

\subsection{Task-Specific Token Decoupling}
In industrial ranking systems, models are commonly trained under multiple optimization objectives~\cite{majiaqi2018MMoE}, which share a common input feature space while imposing heterogeneous and sometimes conflicting requirements on the learned representations.
This shared optimization setting leads to representation collapse under multi-task supervision, where heterogeneous signals compete within a unified latent space. This competition causes dominant tasks to compress the representation space toward task-specific directions, thereby reducing overall representational expressiveness.
To alleviate this issue, we introduce Task-Specific Token Decoupling, which explicitly separates task-dependent information at the token level. Instead of forcing all objectives to operate on a shared set of tokens, we assign task-specific tokens that evolve independently within the same backbone. Formally, we introduce a set of task-specific tokens where $K$ learnable task tokens $\{\mathbf{x}_{task}^{(k)}\}_{k=1}^K$ are introduced and participate in input representation. These tokens are jointly processed with shared tokens but maintain task-specific representations during feature mixing.
Furthermore, the task-specific token is fed directly into its corresponding task tower along with the final output $\mathbf{H}'_L$ of models
$$y^{(k)} = \text{Tower}^{(k)}(\mathbf{x}_{task}^{(k)},  \text{Pool} (\mathbf{H}'_L))$$
This design reduces interference among heterogeneous learning signals and alleviates the compression of the shared representation space into task-biased subspaces. As a result, it preserves higher representational capacity and enables more balanced utilization of the latent space across multiple objectives.

\section{Experiments}

\subsection{Experimental Setting}

\subsubsection{DataSet}
We evaluate the proposed framework on a real-world large-scale industrial dataset from WeChat Video Accounts Ads, comprising 20 million daily samples with over 1,200 sparse features. The dataset spans from July 2024 to March 2026, covering user interactions in a production system.
For model training, we use historical logged data collected from the production system, while evaluation is conducted in an online setting where predictions are computed in real time to reflect practical deployment conditions.
Our evaluation focuses on the Click Conversion Rate (CVR) prediction task, which aggregates 32 online optimization objectives from the production system. Each sub-task corresponds to a distinct business target and is modeled within a unified multi-task learning framework.
The primary baseline is RankMixer~\cite{zhu2025rankmixer}, a sota MetaFormer-based architecture for industrial ranking that achieves an optimal trade-off between computational efficiency and model performance. To ensure a fair comparison, all experiments maintain a consistent two-layer backbone configuration. 

\subsubsection{Evaluation Metrics}
To capture the performance in a dynamic industrial environment, we employ Realtime AUC as our primary metric. Realtime AUC is calculated within short, continuous time windows to monitor the model's discriminative power against evolving data distributions.

To quantify representation quality, we monitor the Effective Rank (erank) of the hidden states produced by each block, as defined in Section 2.2. Specifically, for a given layer output, we collect the hidden representation tensor $\mathcal{H} \in \mathbb{R}^{B \times T \times D}$, where $B$ is the batch size. The effective rank is computed on each sample-wise token representation matrix $\mathbf{H}_b \in \mathbb{R}^{T \times D}$ within the mini-batch. The final metric is reported as the empirical mean across samples and evaluated separately at different network depths to reflect layer-wise representation evolution.
To further clarify the analysis, we denote the intermediate representations after the Token Mixer and the Feed-Forward Network (FFN) within each block as ${TM}$ and ${FFN}$, respectively, where both include residual connections.

\subsection{Results}
To evaluate the effectiveness of RankUp, we conduct ablation studies on key components of the proposed architecture within the WeChat Video Accounts advertising system. We focus on three representative tasks, as they are among the most resource-intensive and high-traffic objectives in the production system, covering diverse business scenarios: Task Order, Task Book, and Add Service.

As shown in Table~\ref{tab:overall_imp_auc}, our framework consistently improves Real-time AUC over the baseline backbone across all evaluated tasks.
Each variant independently introduces a specific component of RankUp and demonstrates consistent gains over the baseline. In particular, the variant with Global Token and Multi-embedding yields the largest improvement, suggesting that enriching the initial representation space and incorporating global contextual information are beneficial for downstream prediction performance.

\begin{table}[th]
    \centering
    \caption{Improvement of Realtime AUC over top-3 tasks}
    \begin{tabular}{c|c|c|c}
    \hline
         & Order & Book & Add Service \\ \hline
    Randomized Permutation Split     & +0.06\% & +0.06\% & +0.08\%\\ 
    w/ Global Token + Multi-Emb & +0.21\% & +0.18\% & +0.13\% \\ 
    w/ Cross Embedding & +0.22\% & +0.10\% & +0.03\% \\ 
    w/ Task Token & +0.09\% & +0.02\% & +0.02\%\\  \hline
    RankUp & +0.41\% & +0.23\% & +0.25\% \\ \hline

    \end{tabular}
    \label{tab:overall_imp_auc}
\end{table}

\subsection{Effect of Split Strategy}
To validate the efficacy of our Randomized Permutation Splitting, we conduct an analysis of the token statistical properties.

\subsubsection{Token Independence via Mutual Information}
We evaluate the statistical independence among tokens by comparing the pairwise Mutual Information (MI)~\cite{cover1999elements} under semantic grouping versus our randomized strategy. Concretely, given a training batch with embedding tensor $\mathbf{E} \in \mathbb{R}^{B \times T \times D}$, we flatten the $B \times T$ embeddings and assign each to one of $K$ clusters via $k$-means, yielding a discrete cluster label $c_i^{(b)} \in \{1, \dots, K\}$ for the $i$-th token in the $b$-th sample. These cluster assignments serve as discrete proxies for the latent states of each token. We then compute a pairwise MI matrix $\mathbf{M} \in \mathbb{R}^{T \times T}$, where each entry is defined as: \begin{equation} M_{ij} = \sum_{a=1}^{K} \sum_{b=1}^{K} p(a, b) \log \frac{p(a, b)}{p_i(a)\, p_j(b)}, \end{equation} where $p(a, b)$ denotes the joint probability that token $i$ is assigned to cluster $a$ and token $j$ to cluster $b$, while $p_i(a)$ and $p_j(b)$ are the corresponding marginal probability. A higher $M_{ij}$ indicates greater statistical redundancy between token $i$ and $j$. To isolate the effect of the randomized strategy, we define the MI Difference Matrix as: \begin{equation} \Delta \mathbf{M} = \mathbf{M}_{\text{Randomized}} - \mathbf{M}_{\text{Semantic}}. \end{equation} Negative entries in $\Delta \mathbf{M}$ signify that randomized assignment reduces inter-token redundancy over semantic grouping, resulting in improved feature decoupling.

\begin{figure}[htbp]
    \centering
    \begin{minipage}{0.235\textwidth}
        \centering
        \includegraphics[width=\textwidth]{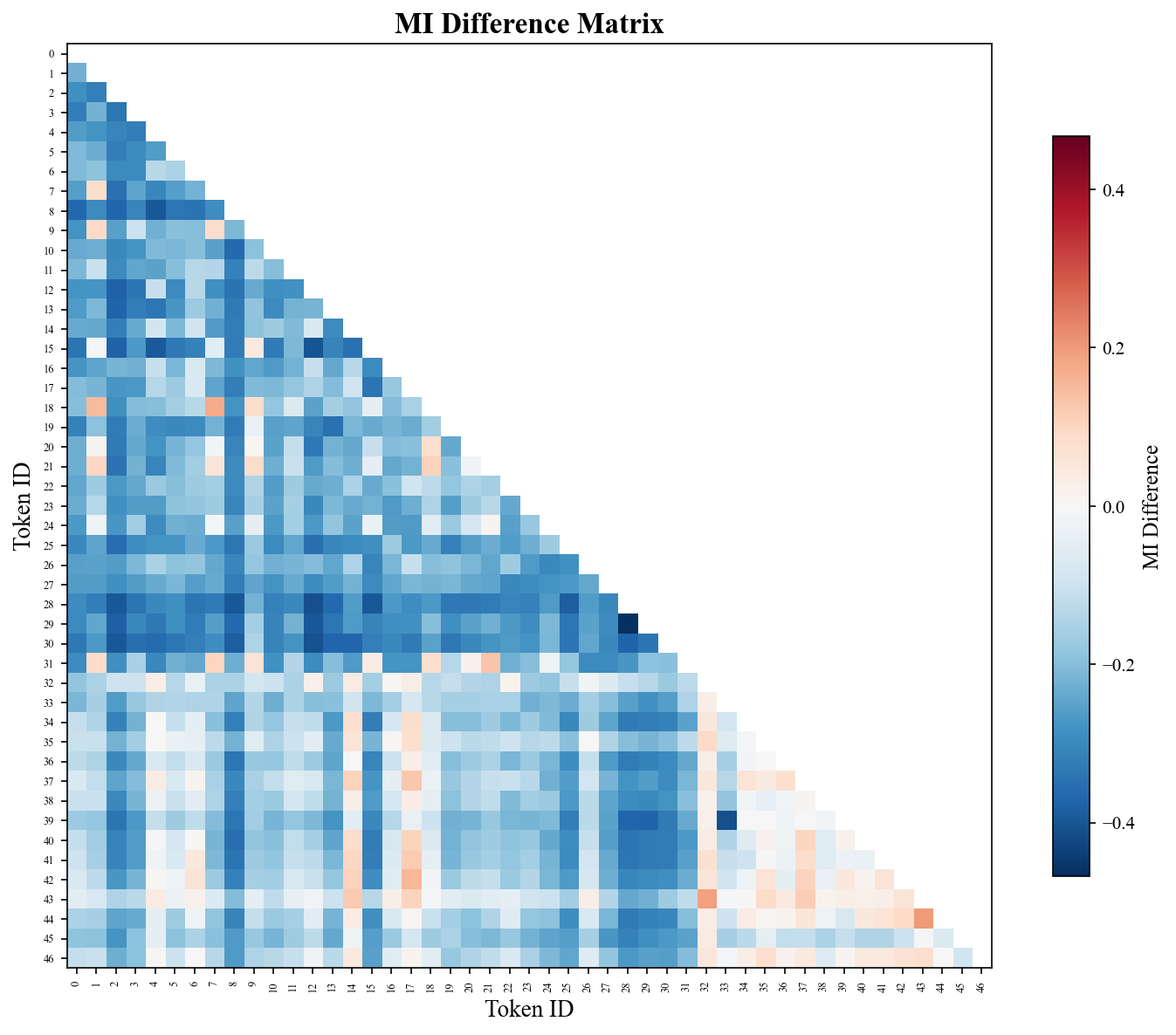}
        \small (a) 48 clusters
        \label{fig:mi_k48}
    \end{minipage}
    \hfill
    \begin{minipage}{0.235\textwidth}
        \centering
        \includegraphics[width=\textwidth]{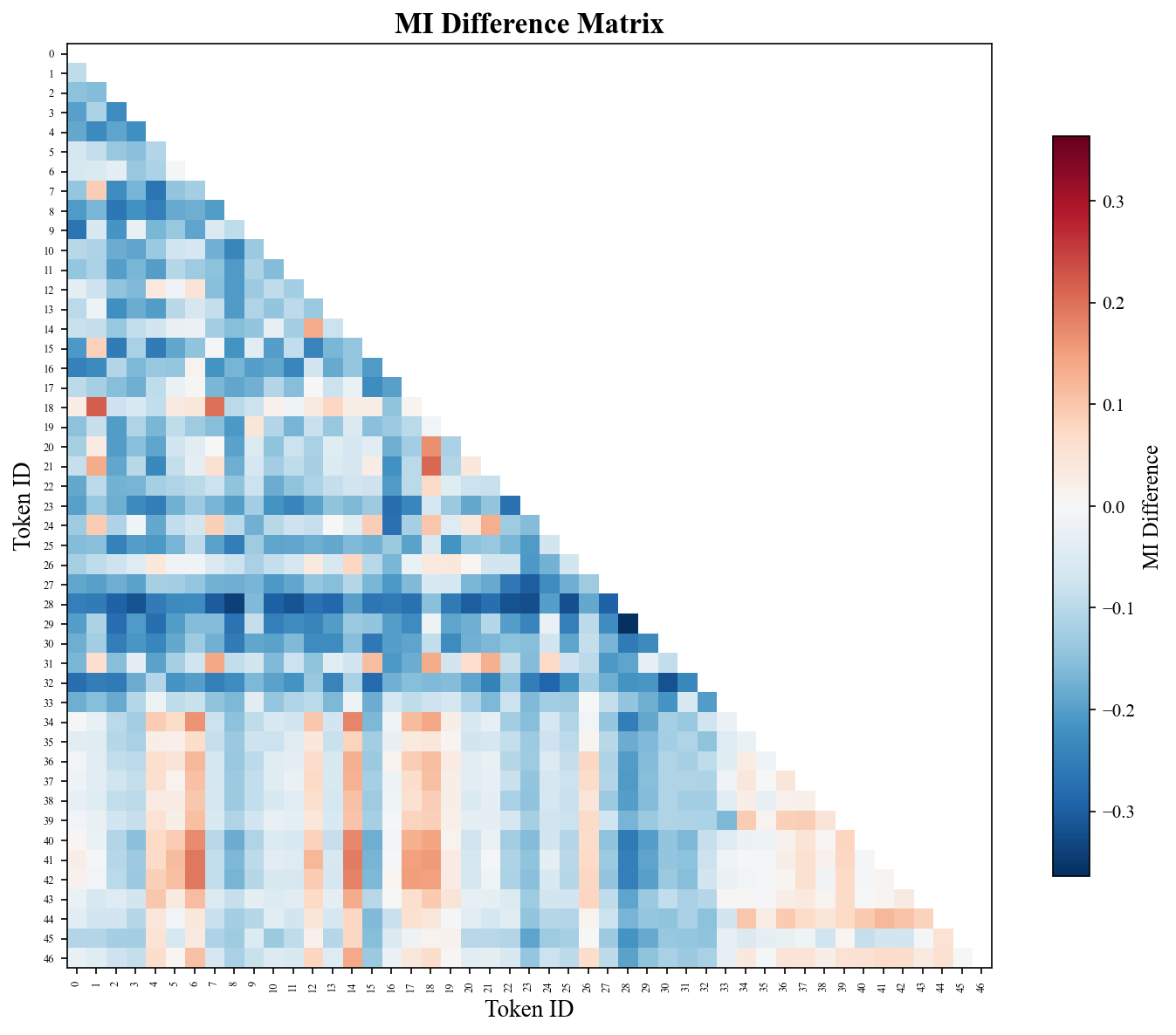}
        \small (b) 64 clusters
        \label{fig:mi_k64}
    \end{minipage}
    \caption{MI difference matrices ($\mathbf{M}_{\text{Randomized}} - \mathbf{M}_{\text{Semantic}}$) of the lower triangle with $K{=}48$ (left) and $K{=}64$ (right) clusters. Blue entries indicate lower MI under the randomized strategy, demonstrating reduced inter-token redundancy. The pattern is consistent across different choices of $K$.}
    \label{fig:mi_diff}
\end{figure}

As shown in Figure~\ref{fig:mi_diff}(a), where embeddings are clustered into 48 centroids, the MI between sparse tokens (Token ID 0–31) is consistently lower under the randomized strategy than under semantic grouping, indicating that our approach yields more statistically independent token representations. For non-sparse tokens (Token ID 32–46), which are constructed identically across both methods, the pairwise MI remains largely unchanged, as expected. Notably, the cross-group MI between sparse and non-sparse tokens is also substantially reduced under the randomized strategy, further demonstrating its effectiveness in disentangling token-level dependencies across heterogeneous feature groups.

We further validate the robustness of this observation by increasing the number of clusters to 
$K=64$. As shown in Figure~\ref{fig:mi_diff}(b), the MI difference matrix exhibits a consistent pattern, confirming that the above findings are not sensitive to the choice of $K$.

\subsubsection{Effective Rank}
\begin{figure}[ht]
    \centering
    \includegraphics[width=\linewidth]{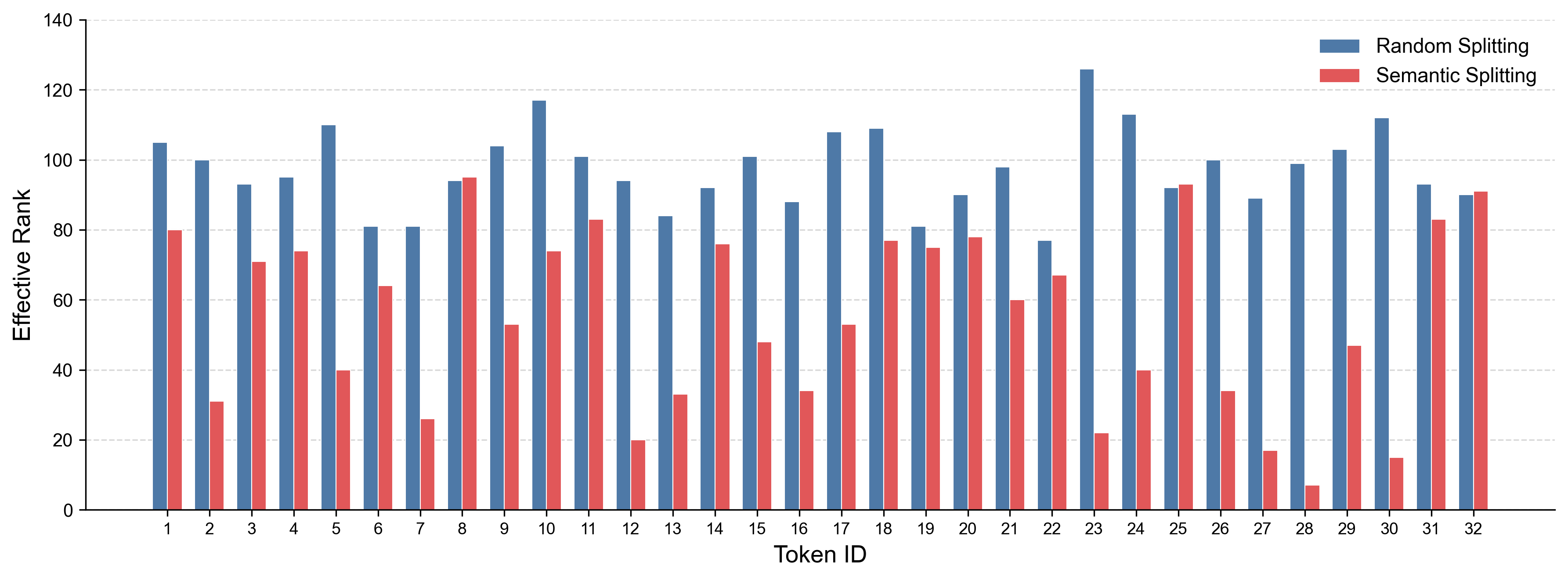}
    \caption{Effective rank comparison of token embeddings under randomized splitting and semantic splitting. The randomized strategy yields consistently higher and more uniform effective ranks across all 32 sparse tokens, while semantic grouping suffers from rank collapse on tokens dominated by long-tail features.}
    \label{fig:token_effective_rank}
\end{figure}

We further examine the effective rank of each token's embedding matrix to assess the diversity of learned representations. A higher effective rank indicates that the embedding spans a higher-dimensional subspace, reflecting greater representational capacity.
As shown in Figure~\ref{fig:token_effective_rank}, the effective rank of tokens under the randomized strategy is consistently higher and more uniform across all 32 sparse tokens compared to semantic grouping. Under semantic grouping, tokens that aggregate long-tail features—which individually have low cardinality and limited variation—tend to exhibit significantly lower effective ranks (e.g., Token 12, 29, and 31 fall below 20). This is because semantically similar long-tail features are grouped into the same token, resulting in a low-rank embedding subspace. In contrast, the randomized strategy disperses such features across different tokens, effectively mitigating the rank collapse caused by long-tail concentration and ensuring that each token maintains a rich and diverse representation.

\subsection{Analysis of Effective Rank Dynamics}
\begin{figure}
    \centering
    \includegraphics[width=\linewidth]{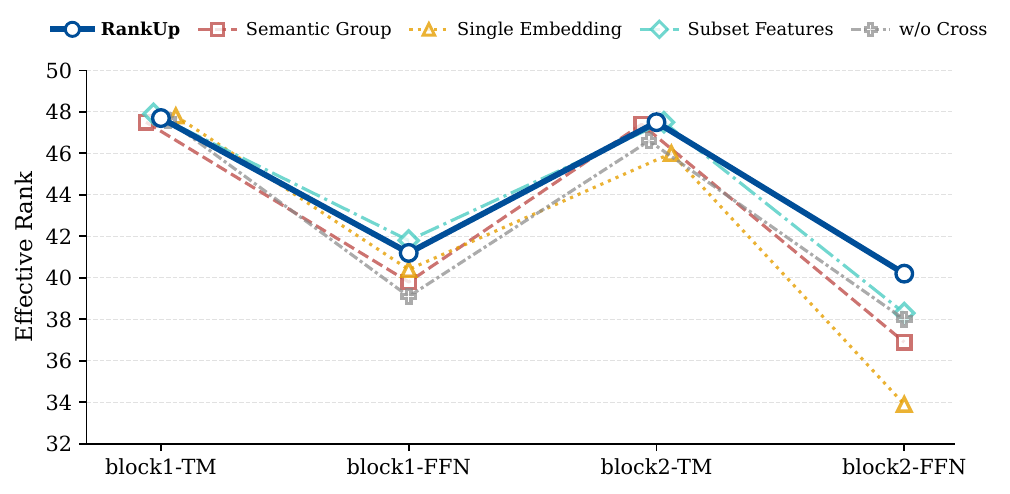}
    \caption{Layer-wise Effective Rank Evolution under RankUp Ablations. This figure illustrates the distinct impact of different RankUp components on layer-wise representation capacity.}
    % It is observed that erank is generally lower in FFN layers than in TM layers, while the ''Rank-Up" method effectively mitigates the rank collapse at block2-FFN.}
    \label{fig:erank_comparison}
\end{figure}

We further investigate how different components of RankUp influence representation quality across layers by analyzing the evolution of effective rank at different layers and different modules, as shown in Figure~\ref{fig:erank_comparison}.

Overall, we observe a consistent pattern of representation degradation in deeper layers, particularly within the FFN stages, where the effective rank decreases significantly. This observation is consistent with recent analyses~\cite{guo2026rankelastor}, which report that deeper architectures often fail to exhibit monotonic rank growth, instead showing degraded representation capacity in later layers.
Meanwhile, RankUp variants maintain higher effective rank across both Token Mixer and Feed-Forward Network layers, indicating improved preservation of representation diversity throughout depth.

To further understand the contribution of each component, we conduct an ablation study by independently modifying key modules in RankUp and analyzing their effects on effective rank dynamics.
The results show a clear separation of roles across components. The Multi-embedding paradigm primarily improves early-stage representation diversity by mapping identical raw features into independent embedding subspaces, preventing early compression and promoting a more isotropic latent space. The Global Token mechanism stabilizes deep-layer dynamics by enabling global feature aggregation, which mitigates information loss observed in subset-based variants where effective rank degrades over depth (e.g., from 40 to 38). The Cross Pre-trained Embedding contributes mainly at the initial stage, where removing it leads to a sharp drop in effective rank at Block 0, indicating that external semantic priors are crucial for forming a rich initialization space. Finally, Randomized Permutation Splitting and cross-embedding interaction further enhance token-level diversity by reducing feature correlation and strengthening cross-feature interactions, leading to smoother rank propagation across layers.
Overall, these components act in a complementary manner, addressing representation collapse from different perspectives.

\subsection{Effect of Task Tokens}
To evaluate the contribution of task-specific tokens to representation learning, we measure the alignment between learned representations and downstream tasks using Mutual Information (MI). Since hidden representations lie in a continuous high-dimensional space, we adopt a discretization-based estimation strategy by flattening token representations and partitioning the latent space into $K$ discrete regions via K-means clustering.
The Mutual Information between the resulting cluster assignments $Z$ and the ground-truth labels $Y \in \{0,1\}$ is computed as
$$
I(Z; Y) = \sum_{z, y} P(z, y) \log \frac{P(z, y)}{P(z)P(y)},
$$
where the joint and marginal probabilities are estimated using empirical frequencies over the evaluation set.
A higher MI indicates stronger alignment between the learned representation structure and downstream task labels, suggesting that the model preserves more task-relevant information while suppressing irrelevant noise.

\begin{figure}
    \centering
    \includegraphics[width=\linewidth]{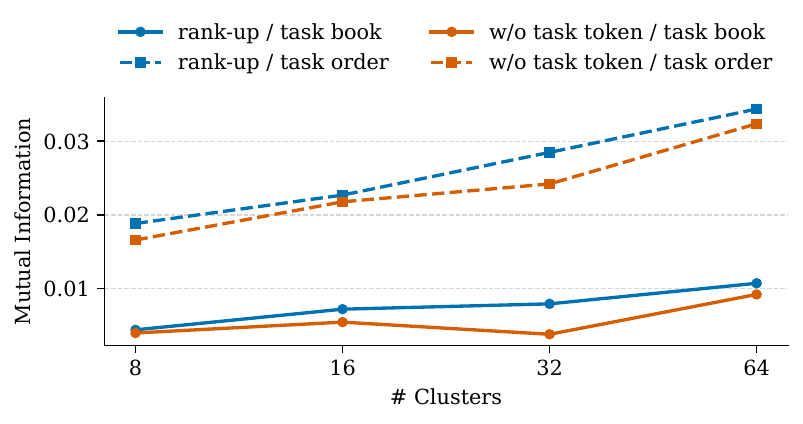}
    \caption{Mutual Information across Different Cluster Granularities. Task-specific tokens consistently improve representation-task alignment, with larger gains observed under finer-grained clustering.}
    \label{fig:mi_clusters_tasks}
\end{figure}

As shown in Figure~\ref{fig:mi_clusters_tasks}, RankUp with task-specific tokens consistently achieves higher Mutual Information than the variant without task tokens across both \textit{book} and \textit{order} tasks. This improvement remains stable under different clustering granularities, ranging from 8 to 64 clusters, indicating that task tokens help preserve more task-relevant information in learned representations.

Notably, as the number of clusters increases, the MI gap between RankUp and the baseline becomes progressively larger. Since finer-grained clustering reflects a more detailed partitioning of the latent space, this trend suggests that task tokens improve not only coarse-grained task awareness but also the model's ability to capture subtle task-specific structures.
Overall, these results demonstrate that task-specific tokens encourage the model to organize latent representations in a task-aware manner, leading to better disentanglement of predictive signals from irrelevant noise.

\section{Online Performance}
RankUp has been \textbf{fully deployed} in Tencent Weixin's production advertising systems, including pcvr models for Weixin Video Accounts, Official Accounts and Moments.

\subsection{Deployment Details}
We evaluate RankUp in a large-scale industrial advertising system for Conversion Rate (CVR) prediction. All model variants are trained from scratch under the same training protocol without initializing from historical checkpoints. The training data consists of 18 months of large-scale user--ad interaction logs collected from real-world production traffic.

Our framework is deployed across three major high-traffic advertising scenarios: Video Accounts, Official Accounts and Moments. To capture diverse user behaviors in these complex environments, each model is designed as a multi-task predictor jointly optimizing 32 prediction tasks within the advertising ecosystem.

The input layer operates over a high-dimensional sparse feature space containing more than 1,000 feature fields, including domain-specific categorical features, sequential behavior tokens extracted from sequence encoders, and multiple pre-trained semantic embeddings. This design enables the model to jointly exploit sparse behavioral patterns and dense semantic signals during inference.

For deployment, we adopt a 2-layer MetaFormer backbone as the core architecture. To support industrial-scale representation learning, the model size is scaled up by an order of magnitude compared to previous production systems, increasing from approximately 10M to 100M parameters per scenario. All experiments are conducted against the current production ranking system, such as the rankmixer as a submodule in Weixin Official Accounts. The deployed model operates with a batch size of 300, requiring approximately 70 GFLOPs per batch. Notably, RankUp achieves a Model FLOPs Utilization (MFU) of 23\%, indicating improved hardware efficiency despite the increased model capacity. Despite the increased computational complexity introduced by RankUp, the system remains within strict industrial latency constraints and is fully compatible with real-time serving requirements.

\subsection{Online Deployment Results}
To evaluate the effectiveness of RankUp in real-world industrial environments, we conducted large-scale online A/B testing on \textbf{20\%} of production traffic across three major advertising scenarios: Weixin Video Accounts, Weixin Official Accounts and Weixin Moments. All experiments were conducted against the existing production ranking system under identical serving conditions. Each A/B experiment was conducted continuously for 14 days to ensure statistical stability. All reported improvements are statistically significant under repeated daily measurements. We report key business metrics including Realtime AUC, CTCVR, and GMV to comprehensively assess ranking quality, advertising efficiency, and revenue impact.

As summarized in Table~\ref{tab:online_overall_res}, RankUp consistently improves online ranking performance across all deployment scenarios. In particular, the proposed model achieves up to \textbf{0.367\%} AUC gain in Weixin Video Accounts, demonstrating stronger ranking discrimination ability in real production settings. More importantly, these ranking improvements translate directly into business value, yielding \textbf{2.21\%--4.81\%} GMV lift across all scenarios while maintaining healthy advertising cost efficiency.

Given the scale of our advertising platform, even marginal improvements in online metrics produce significant commercial impact. The observed gains correspond to an estimated increase of \textbf{hundreds of millions of dollars} in annual revenue. Based on these robust results, RankUp has been \textbf{fully deployed} to serve \textbf{100\%} of live production traffic.
\begin{table}[h]
    \centering
    \caption{Online Performance Lift}
    \begin{tabular}{lccc}
        \toprule
        Scenario & $\Delta$ AUC & CTCVR & GMV  \\
        \midrule
        Weixin Video Accounts & +0.367\% & +1.41\% & +3.41\%  \\
        Weixin Official Accounts & +0.331\% & +0.21\%  & +4.81\% \\
        Weixin Moments & +0.269\% & +0.87\% & +2.12\%  \\
        \bottomrule
    \end{tabular}
    \label{tab:online_overall_res}
\end{table}

\begin{table}[th]
    \centering
    \caption{Online Performance Lift for New Ads}
    \begin{tabular}{lc}
        \toprule
        Scenario &  GMV \\
        \midrule
        Weixin Video Accounts & +5.83\%  \\
        Weixin Official Accounts & +9.67\% \\
        Weixin Moments & +2.84\%  \\
        \bottomrule
    \end{tabular}
    \label{tab:online_new_ads}
\end{table}

Table~\ref{tab:online_new_ads} further highlights the effectiveness of RankUp in handling newly launched advertisements, which represent a challenging cold-start scenario in industrial recommendation systems due to extreme data sparsity. By leveraging richer representation learning and stronger feature interaction modeling, RankUp generates more expressive representations even when behavioral supervision is limited.
Specifically, for first-day performance of new advertisements in Weixin Official Accounts, RankUp achieves a remarkable \textbf{9.67\%} GMV increase. These results indicate that RankUp substantially improves the discoverability of high-quality new advertisements, enabling more effective exploration and monetization during the cold-start stage.

\begin{table}[th]
    \centering
    \caption{Online Performance Lift for Order Task}
    \begin{tabular}{lc}
        \toprule
        Scenario &  GMV   \\
        \midrule
        Weixin Video Accounts &  5.18\%  \\
        Weixin Official Accounts  & 7.18\%  \\
        Weixin Moments & 4.79\%  \\
        \bottomrule
    \end{tabular}
    \label{tab:online_order_task}
\end{table}

The effectiveness of RankUp is further validated on the Order Task, which constitutes the largest proportion of advertising consumption in our platform and serves as the primary optimization objective for conversion quality. As shown in Table~\ref{tab:online_order_task}, RankUp significantly improves this critical task, yielding \textbf{7.18\%} GMV lift in Weixin Official Accounts.
These results suggest that RankUp effectively enhances representation quality for high-value conversion objectives, allowing the ranking model to better capture and preserve subtle purchase intent signals. Given the dominant business importance of the Order Task, such gains further demonstrate the practical value of RankUp in improving large-scale industrial recommendation systems.
\section{Related Work}
\subsection{Dense Scaling Laws in Rec}
Alongside the continued scaling of embedding tables, a parallel thread of research has focused on upscaling the dense interaction network—a paradigm we term dense scaling—to unlock further performance gains. Early attempts at deepening or widening interaction layers yielded diminishing returns due to CPU-era architectural constraints. AutoInt~\cite{songweiping2019AutoInt} pioneered the use of multi-head self-attention to explicitly model high-order feature interactions, though its homogeneous treatment of features limited expressiveness. Hiformer~\cite{gui2023hiformer} improved upon this by introducing heterogeneous attention with feature-specific projections, enabling more nuanced interactions. Wukong~\cite{zhang2024wukong} provided the first empirical demonstration of a clear scaling law within the dense component, showing that stacked Factorization Machines could scale quality across two orders of magnitude in computational complexity. A key challenge emerged in balancing sequential behavior modeling with dense feature interaction. InterFormer~\cite{zeng2025interformer} addressed this through bidirectional information flow between non-sequential and sequential features, preventing early summarization and information loss. RankMixer~\cite{zhu2025rankmixer} introduced a hardware-aware architecture with token mixing and per-token FFNs, dramatically improving Model Flops Utilization (MFU) from 4.5\% to 45\% and enabling 100× parameter scaling without increased latency. MixFormer~\cite{huang2026mixformer} unified sequence and dense modeling within a single parameter space, resolving the co-scaling conflict between the two components. Collectively, these works establish the architectural foundations for the scaling laws we investigate in this paper.

\subsection{Representation Collapse in 
Neural Networks and Recommender Systems}
Representation collapse—where learned embeddings concentrate in a low-dimensional subspace and fail to preserve discriminative information across semantically distinct samples—has been widely observed in representation learning paradigms~\cite{balestriero2023cookbook,lijing2021understanding_dimension_collapse_ssl}. This phenomenon is often characterized by the degeneration of embedding diversity, where informative signals are restricted to a few principal directions, leading to a reduced effective rank of the representation matrix and impaired generalization.

Representation collapse has been extensively studied in self supervised and contrastive learning~\cite{hua2021feature,tian2021understanding_ssl,lijing2021understanding_dimension_collapse_ssl,hua2021feature}, which often stems from an imbalance between alignment (bringing similar pairs together) and uniformity (spreading all embeddings across the hypersphere)~\cite{UnderstandingCLfromAlignmentAndUniformity2020}. Without explicit regularization, models may converge to trivial solutions where embeddings become rank-deficient or even constant~\cite{bardes2021vicreg}. Recent advances in Large Language Models (LLMs) have also identified similar patterns; for instance, the effective rank of attention outputs can fluctuate across layers~\cite{wang2020linformer,tay2021synthesizer}, with MLP blocks and residual connections playing a critical role in re-ranking the representation space to prevent collapse~\cite{wang2025attention}.

Representation collapse is further exacerbated by the inherent properties of industrial-scale data in recommender systems, such as extreme sparsity and power-law distributions \cite{pan2024ads,guo2023embedding,ardalani2022understanding}. 
In particular, the long-tailed distribution of item frequencies often leads to a dominance of head items in the embedding space, while tail items receive insufficient updates. 
To mitigate these issues, prior studies have proposed a variety of strategies. One line of work introduces regularization terms or auxiliary losses to encourage embedding dispersion, such as uniformity-promoting objectives or orthogonality constraints~\cite{wang2022towards,loveland2025understanding,cai2024popularity}. As dense models continue to scale in industrial settings, ensuring the effectiveness and robustness of learned representations has become increasingly crucial for maintaining system performance.

\section{Conclusion}
In this work, we revisit the scaling behavior of modern recommendation architectures from the perspective of representation capacity. While prior studies have validated the empirical scaling laws of MetaFormer-based recommenders, we show that parameter growth does not necessarily translate into proportional gains in expressive power, as deeper architectures often suffer from representation collapse and limited effective-rank improvement. To address this limitation, we propose \textbf{RankUp}, a novel architecture designed to mitigate representation collapse by enhancing representation diversity through randomized permutation splitting, multi-embedding representations, global token integration, crossed pre-trained embedding tokens, and task-specific token decoupling. Extensive experiments demonstrate that RankUp consistently improves effective rank and recommendation performance across large-scale industrial datasets. Moreover, RankUp has been successfully deployed in production across multiple Tencent advertising scenarios, delivering significant GMV gains in real-world applications. Our findings highlight that scaling recommendation models should emphasize not only parameter growth, but also the effective utilization of representation space.

\newpage
\bibliographystyle{ACM-Reference-Format}
\bibliography{7.reference}

\end{document}